\documentclass{article}

\usepackage[preprint]{neurips_2020}

\usepackage[utf8]{inputenc} % allow utf-8 input
\usepackage[T1]{fontenc}    % use 8-bit T1 fonts
\usepackage{hyperref}       % hyperlinks
\usepackage{url}            % simple URL typesetting
\usepackage{booktabs}       % professional-quality tables
\usepackage{amsfonts}       % blackboard math symbols
\usepackage{nicefrac}       % compact symbols for 1/2, etc.
\usepackage{microtype}      % microtypography
\usepackage{amsmath}
\usepackage{float}
\usepackage{multicol}
\usepackage{graphicx}
\usepackage{caption}
\usepackage{color}
\usepackage{wrapfig}
\usepackage[square,numbers]{natbib}

\title{Approximating Ground State Energies and Wave Functions of Physical Systems with Neural Networks}

\author{
  Cesar Lema \thanks{undergraduate student}\\
  Department of Physics\\
  New York University\\
  \texttt{cl4393@nyu.edu} \\
  \And
  Anna Choromanska \\
  Department of Electrical and Computer Engineering\\
  New York University\\
  \texttt{ac5455@nyu.edu} \\
}

\begin{document}

\maketitle

%\vspace{-0.2in}
\begin{abstract}
%\vspace{-0.05in}
Quantum theory has been remarkably successful in providing an understanding of physical systems at foundational scales. Solving the Schr\"{o}dinger equation provides full knowledge of all dynamical quantities of the physical system. 
However closed form solutions to this equation are only available for a few systems and approximation methods are typically used to find solutions. In this paper we address the problem of solving the time independent Schr\"{o}dinger equation for the ground state solution of physical systems. We propose using end-to-end deep learning approach in a variational optimization scheme for approximating the ground state energies and wave functions of these systems. A neural network realizes a universal trial wave function and is trained in an unsupervised learning framework by optimizing the expectation value of the Hamiltonian of a physical system. The proposed approach is evaluated on physical systems consisting of a particle in a box with and without a perturbation. We demonstrate that our approach obtains approximations of ground state energies and wave functions that are highly accurate, which makes it a potentially plausible candidate for solving more complex physical systems for which analytical solutions are beyond reach.
\end{abstract}

\vspace{-0.1in}
\section{Introduction}
\label{heading_a}
\vspace{-0.05in} 
In the field of quantum mechanics the state of a physical system is given by a wave function. The wave function is mathematically represented in Dirac notation as an abstract vector $|\Psi \rangle$ in a complex vector space~\cite{GriffithsBook,DiracBook}. In contrast, the state of a physical system in classical mechanics is given by a point in a phase space~\cite{TaylorBook}. The usefulness of the wave function is confirmed by Born's statistical interpretation~\cite{Born1926-BORZQD}, which  has been accepted as a postulate in foundations of quantum
mechanics and interprets the modulus square of a wave function as the probability of a measurement of an observable at any given time (i.e., the probability for a specific result in the measurement of an observable $X$ at any given time $t$ is given by the modulus square of the wave function in the eigenbasis of $X$ i.e. $|\langle x | \Psi(t)\rangle|^2$). Observables are quantities of a physical system that can be measured and in quantum theory are mathematically represented by Hermitian operators~\cite{DiracBook}. As an example, the knowledge of the wave function allows computing the probability of finding a particle at any given position or a particle having a specific momentum.

Access to the wave function provides full knowledge of all dynamical quantities of a physical system.
In quantum mechanics, dynamical variables, such as the kinetic or angular momentum of a system, are represented by a combination of position and momentum operators. Their values are computed as expectation values of the corresponding operator given a wave function. This is analogous to classical mechanics, where all dynamical quantities can be represented as a function of position and momentum. Thus, the wave function ultimately enables to analyze the behavior of physical systems, including those that cannot be easily studied experimentally.  

The energy states of a non-relativistic quantum systems~\cite{LandauBook, PielaBook}, which involve particles with speeds much less than the speed of light, are found by solving the time independent Schr\"{o}dinger equation (TISE) given as 
\begin{equation}
H|\Psi_n\rangle = E_n|\Psi_n\rangle,
\end{equation}
where $H$ is the Hamiltonian of a quantum system  (Hamiltonian is a Hermitian operator and is equal to the sum of the kinetic energy $T$ and potential energy $V$ operators), and $\Psi_n$ and $E_n$ are the $n^{th}$ eigen function and eigen value of $H$ respectively. TISE highlights that the variables characterizing quantum physical systems, such as momentum, spin, energy, etc., are quantized. TISE is an eigenvalue problem for the Hamiltonian operator $H$ and its solution gives the eigenbasis and stationary states for the energy of a system. TISE is presently the major basic tool of many branches of modern physics~\cite{LaloeBook} as well as chemistry~\cite{hermann2020deep}. For many physical systems, such as electron configuration of atoms, molecular and strongly interacting many body systems, there are no closed form solutions to TISE~\cite{PielaBook,hermann2020deep,carleo}. This work offers an accurate numerical way of solving this equation and presents a deep learning based optimization approach for approximating ground state solutions to TISE.

\vspace{-0.05in}
\section{Prior work}
\vspace{-0.05in}

Solving TISE typically requires numerical approximations. Among approximation methods, the most popular are Hartree–Fock, configuration interaction, and coupled cluster~\cite{CoesterPaper} methods, conveniently reviewed in~\cite{PielaBook}, as well as multideterminant quantum Monte Carlo technique~\cite{MoralesPaper}. The variational principle is the foundation for some commonly used methods~\cite{PielaBook} and provides a simple and flexible scheme for finding the ground state solution to TISE . The variational principle gives an upper bound for the ground state energy $E_{\text{gs}}$ of a given (Hermitian) Hamiltonian operator $H$. The expectation value of the Hamiltonian $\langle H \rangle = \langle \Psi |H| \Psi \rangle$ (Eq. \ref{eq:2}) with respect to an arbitrary wave function 
$|\Psi\rangle$ (from a complex Hilbert space) is the energy $\langle E \rangle$ of the system in the state $|\Psi\rangle$. By definition $E_{\text{gs}}$ (the lowest eigenvalue of $H$) is the lowest possible energy of the Hamiltonian, therefore $\langle E \rangle \geq E_{gs}$. The choice of the trial wave function $|\Psi\rangle$ determines the tightness of the upper bound on $E_{gs}$ for a given Hamiltonian and consequently its quality. Limited prior knowledge of the wave function, furthermore, affects the accuracy of this method. 

Deep learning models have shown incredible aptitude in solving difficult learning problems and could be used to represent even highly complicated trial wave functions. Therefore deep learning can be perceived as a plausible candidate method for recovering the wave function that leads to tight variational bound and accurate approximation of the ground state of a physical system. Integrating artificial neural networks in a variational optimization method to solve for the ground state solution of TISE was explored only recently. For spin lattice many-body systems a restricted Boltzmann machine was used to realize the wave function ansatz in the variational Monte Carlo (VMC) method~\cite{carleo}. For the electronic Schr\"{o}dinger equation, Deep Neural Networks (DNN) integrated into a wave function ansatz was also explored in the VMC setting ~\cite{HanPaper, PfauPaper, hermann2020deep}. More recently, \cite{hermann2020deep} presented a deep learning wave function ansatz that achieved state-of-the-art results for solutions of the electronic Schr\"{o}dinger equation (authors also conveniently summarize the existing methods for solving TISE problem in an online video~\cite{Video}).
For simple Hamiltonians, the radial basis function network~\cite{peiyuan} was used to realize the wave function ansatz also in the VMC method. \\
\\ 
For arbitrary Hamiltonians in the VMC setting, the optimal wave function ansatz are not generally known and there exists a trade off between complexity of modeling the wave function and its accuracy~\cite{Kim_2018, PangBook, becca_sorella_2017}. The work presented here shows how to learn the wave function approximation in the end-to-end deep learning framework. We differ from the aforementioned techniques in a couple of ways: 1) we do not use the fixed expression for the wave function ansatz but instead model it directly via DNN, 2) our method is not constrained to a specific physical system, 3) our technique is simple and does not require sampling steps as opposed to VMC. Due to the high expressiveness of neural networks and the simplicity of our approach, our work has the potential to solve TISE for arbitrary and previously inaccessible systems. It also provides a general framework for solving TISE of discrete and continuous systems (i.e. solving TISE in finite and infinite dimensional Hilbert spaces). 

\vspace{-0.05in}
\section{Approach}
\label{heading_b}
\vspace{-0.05in}

We propose to approximate the ground state wave function $\Psi: \mathbb{R}^n\to\mathbb{C}$ ($n$ is the number of particles) that constitutes the ground state solution to the TISE using a DNN. The expectation value of the given Hamiltonian (Eq. \ref{eq:2}) with respect to the wave function approximated with DNN was used as the objective function to train DNN. 
%\textcolor{blue}{By the variational principles we predict the expectation value to be greater than $E_{\text{gs}}$. In a variational scheme we then optimize the NN to give a tighter upper bound.} 
Once DNN approximated the ground state wave function with high fidelity, this expectation value approached the $E_{\text{gs}}$ of the given system. This gives us the ground state solution of TISE we are looking for. Note that using DNN allows operating in the regime where the wave function ansatz is universal and unconstrained. In the context of the variational method, the neural network serves as the trial wave function $|\Psi\rangle$. Using neural networks to  represent the trial wave function allows to explore a rich set of parameterized wave functions. The approach, therefore, heavily hinges on the nature of neural networks as universal function approximators. 

The Hamiltonian consists of operators which cannot be easily applied to a DNN wave function. This makes it difficult to directly compute the expectation value of the Hamiltonian using a DNN ansatz in the variational method. We propose to numerically compute the decomposition of the function approximated by the DNN onto a computational basis (that spans the corresponding Hilbert space) and then compute the expectation value of the Hamiltonian in that computational basis. The expectation value of the energy of a system (Hamiltonian) with respect to the DNN wave function can be computed using an arbitrary computational basis according to the following equation:

\vspace{-0.15in}
\begin{equation}\label{eq:2}
    \langle H \rangle 
    = 
    \frac{
    \langle \Psi| H |\Psi  \rangle} 
    {
    \langle \Psi|\Psi  \rangle} 
    = %%%%%%%%%%%%%%%%%%%%%%%%%%%%%
    \frac{
    \sum_{x,x'}\langle \Psi|x\rangle \langle x| H |x' \rangle \langle x' |\Psi  \rangle} 
    {
    \sum_{x} \langle \Psi| x \rangle \langle x |\Psi  \rangle 
    } 
    = %%%%%%%%%%%%%%%%%%%%%%%%%%%%%
    \frac{
    \sum_{x,x'} \Psi^*(x) H_{x x'} \Psi(x')}
    {
    \sum_{x}  |\Psi(x)|^2 
    },
\end{equation}
\vspace{-0.15in}

where $x, x'$ denote arbitrary basis functions and the decomposition of the neural network trial wave function $|\Psi\rangle$ onto the computational basis is given by $|\Psi  \rangle = \sum_x |x  \rangle \langle x|\Psi  \rangle $. The decomposition coefficients $\Psi(x')=\langle x|\Psi  \rangle$ are generally multidimensional integrals and were estimated using Riemann approximations (Riemann approximations rely on batch computations, i.e., they require wave function values computed over the entire domain which must be finite). The expectation value of the Hamiltonian $\langle x|H|x'\rangle$ computed for two arbitrary basis functions gives $H_{x x'}$.
The computational basis in our experiments was chosen such that calculating $H_{x x'}$ was feasible analytically or computationally. In the experiments we show that despite the computational basis possibly being infinite, we can recover accurate solutions using its finite subset. %Additionally, orthogonal computational basis can be used to efficiently scale the framework.

\vspace{-0.05in}
\section{Experiments}
\label{heading_c}
\vspace{-0.05in}

DNN was trained in an unsupervised setting to approximate the ground state wave function for physical systems given below. We consider the unperturbed and perturbed particle in a box systems since they have known solutions that we can compare our results with. These systems consist of a single one dimensional particle in a real space and therefore the input of the DNN is a scalar value representing the position of the particle in its domain. The output of the network is the approximated value of the ground state wave function. The parameters of the DNN were optimized using batch gradient descent and automatic differentiation tools were used to compute gradients for backpropagation. This was implemented using Pytorch.

The first $N\in \mathbb{Z}^+$ (a hyperparameter of the framework) eigenfunctions of the Hamiltonian of the particle in a box systems were used as the computational basis. Real valued functions were used for $|\Psi\rangle$, but we emphasize that the framework described in this paper would be identical when using complex functions for $|\Psi\rangle$ instead.

%%%%%%%%%%%%%%%%%%%%%%%%%%%%%%%
\subsection{A particle in a box }
\vspace{-0.05in}

A particle in a box consists of a free particle constrained to a domain $x\in\{0,a\}: a\in \mathbb{R}$ by infinite potential outside its domain, i.e., infinite potential well: 
\vspace{-0.05in}
\begin{equation}
    V(x) =
    \begin{cases}
    0 & 0 < x < a
    \\
    \infty & \text{otherwise}
    \end{cases}
\end{equation}
\vspace{-0.15in}

where $a$ is the well width. The Hamiltonian of this system is $H = -\frac{\hbar^2}{2\mu}\frac{d^2}{d x^2}$, where $\mu$ is the particle mass and $\hbar$ is the Planck constant. Its eigenvalues are  $E_n = \frac{n^2 \pi^2 \hbar^2}{2\mu a^2}$ and its eigenbasis is: 
\vspace{-0.05in}
\begin{equation}
    b_n =    
    \begin{cases}
    \sqrt{\frac{2}{a}}
    \sin{(\frac{n \pi}{a} x)}& 0 < x < a
    \\
    0 & \text{otherwise}
    \end{cases}
\end{equation}
\vspace{-0.15in}

where $n\in \mathbb{Z}^+$ indicates the $n^\text{th}$ stationary state. We used the first $N$ eigenfunctions of this system as the computational basis and computed matrix $H_{nm}$ ($m$ also indicates an arbitrary stationary state) as follows: $H_{nm}= \delta_{nm}E_n = \delta_{nm}\frac{n^2 \pi^2 \hbar^2}{2\mu a^2}$. Thus finally, the objective function in this experiment is: 
\begin{align}
        \langle H \rangle
        = %%%%%%%%%%%%%%%
        \frac{
        \sum_{n} |\langle b_n |\Psi \rangle|^2  H_{nn} }
        {
        \sum_{n} |\langle b_n |\Psi \rangle|^2 
        } 
\end{align}
\vspace{-0.15in}

%%%%%%%%%%%%%%%%%%%%%%%%%%%%%%%%%%%%%%%%%
\subsection{A particle in a box with a linear perturbation}
\vspace{-0.05in}

For this experiment a linear potential $V(x) = \alpha x$, where $\alpha\in \mathbb{R}$, was added to the infinite square well potential of the particle in a box system. The Hamiltonian of this system is $H = H^{0} + H' = -\frac{\hbar^2}{2\mu}\frac{d^2}{d x^2} + \alpha x$, where $ H^{0}$ is the Hamiltonian of the unperturbed particle in the box system and $H'$ is the perturbation. The objective function in this experiment was the expectation of the perturbed Hamiltonian given by
\vspace{-0.05in}
\begin{align}
    \langle H \rangle &= 
    \langle H^{0} \rangle 
    + 
    \langle H' \rangle ,
\end{align}
due to the linearity of the expectation (this property was also used to compute the matrix $H_{nm}$ for this experiment). The expectation value of the perturbation is given by:
\vspace{-0.05in}
\begin{align}
    \langle H' \rangle = \alpha \langle x \rangle = \alpha  \frac{\sum_{n} |\langle b_n |\Psi \rangle|^2x_{loc}(b_n)}
    {\sum_n |\langle b_n |\Psi \rangle|^2} 
    , \ 
    x_{loc}(b_n) = \sum_{m} x_{nm} \frac{\langle b_m |\Psi \rangle}{\langle b_n |\Psi \rangle}
\end{align}
\vspace{-0.2in}

The expectation value of the position operator in the computational basis $x_{nm}$ was computed analytically:  
\vspace{-0.1in}
\begin{equation}
    x_{nm} =
    \begin{cases}
    \frac{a}{2} & n=m
    \\
     \frac{a}{\pi^2}
    \left[
    \frac{
    \cos{\left( (n-m)\pi   \right)}
    - 1 }
    {(n-m)^2}
    -
    \frac{
    \cos{\left((n+m)\pi \right)}
    - 1 }
    { (n+m)^2}
    \right] & n\neq m
    \end{cases}.
\end{equation}
\vspace{-0.2in}

%%%%%%%%%%%%%%%%%%%%%%%%%%%%%%%
\subsection{Results} 
\vspace{-0.05in}

\begin{wrapfigure}{r}{8.5cm} %7.5cm
\vspace{-0.4in}
    \centering
    \fbox{\includegraphics[width=0.95\linewidth]{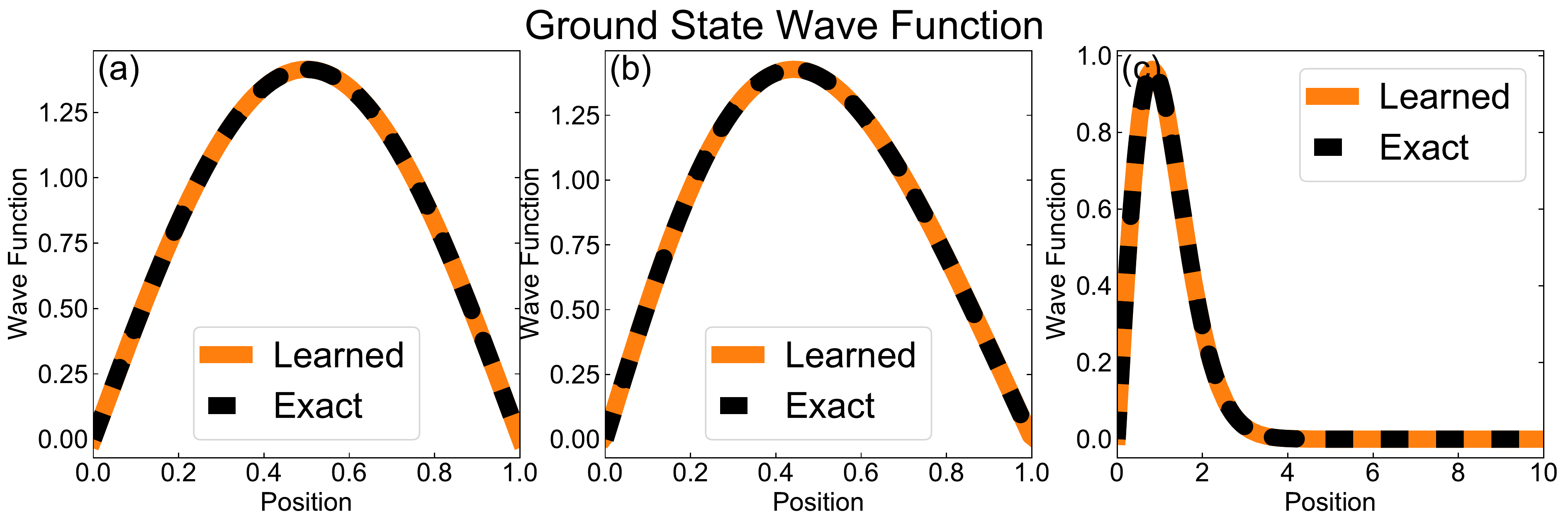}}
    \vspace{-0.05in}
    \caption{\textbf{Normalized ground state wave functions.} Results (solid line) for the unperturbed system (a), perturbed system A (b), and perturbed system B (c) are compared to the exact results (dashed line).}
    \label{fig:wfs}
    \vspace{-0.1in}
\end{wrapfigure}
In our experiment for the unperturbed system (well width $a$=1) we used a DNN with $1$ hidden layer of width $1000$ and ReLU activation functions. For the experiments with perturbed systems (first considered system, system A, had well width $a=1$ and perturbation constant $\alpha=8$ and the second one, system B, had well width $a=10$ and perturbation constant $\alpha=2$) we used DNNs with $2$ hidden layers of width $500$ and $100$ and ReLU activation functions. For all experiments presented $N=100$ basis functions were used. The normalized ground state wave function found by our neural networks is compared to the exact wave function in Figure \ref{fig:wfs}. In Table \ref{GSE} the ground state energies (in natural units) of each system are compared to the exact values as well as the values obtained using the VMC method of~\cite{peiyuan}. The exact values were obtained by solving TISE using a symbolic differential equation solver, i.e., Mathematica.

\begin{wraptable}{r}{7.5cm}
  \vspace{-0.2in}
  \caption{Ground state energies}
  \vspace{-0.1in}
  \label{sample-table}
  \centering
  %\resizebox{0.4\textwidth}{!}{
  \begin{minipage}{7.5cm} % For citation footnotes
  \begin{tabular}{llll}
    \toprule
                &\multicolumn{3}{c}{$E_{\text{ground state}}$} \\
    \cmidrule(r){2-4}
    System      & Computed   & VMC~\cite{peiyuan} & Exact \\
    \midrule
    Unperturbed & 4.93484  & 4.9348  & 4.93480
    \\
    Perturbed A & 8.79510 & 8.7960  & 8.79507
    \\
    Perturbed B & 2.94583 & NA  & 2.94583 \\
    %\footnote{Well width $a=10$ and perturbation constant $\alpha=2$.} & 2.94583 & -  & 2.94583
    \bottomrule
  \end{tabular}
  \end{minipage}
  %}
  \label{GSE}
  \vspace{-0.15in}
\end{wraptable}

Note that the ground state wave functions obtained with our DNN model are practically identical to the analytically obtained target wave functions (Figure \ref{fig:wfs}). Moreover, as can be observed in Table \ref{GSE}, the proposed approach finds highly accurate approximations of the ground state solutions, which are also comparable or better than the approximations obtained with the VMC method of~\cite{peiyuan}. We also accurately approximated system (perturbed system B) that corresponds to a more complicated wave function that was not analyzed in~\cite{peiyuan} The obtained preliminary experiments confirm the plausibility of using DNNs to represent the trial wave function and effectiveness of the proposed variational method. Such approach can be especially useful in case of systems where there exist a limited prior knowledge to motivate a specific choice of the trial wave function. In future works we will generalize our approach to approximate the solution of TISE for more complex physical systems~\cite{hermann2020deep, HanPaper}. 

\vspace{-0.05in}
\section{Conclusion}
\label{heading_d}
\vspace{-0.05in}

In this manuscript a new framework for approximating the ground state solution of the physical system described by a given Hamiltonian is presented. It is based on using a deep neural network to represent a trial wave function in a variational optimization scheme via an end-to-end deep learning framework. The DNN is trained by minimizing the expectation value of the Hamiltonian. Initial experimental results on particle in a box physical systems are presented and confirm the validity of the proposed approach.

\bibliographystyle{unsrt}
\bibliography{mybib}

\end{document}